\documentstyle[12pt,bezier]{article}
\advance\hoffset by -7mm  
\renewcommand{\title}[1]{\null\vspace{22mm}

\noindent{\Large{\bf #1}}\vspace{10mm}

\noindent {\large By }}
\newcommand{\authors}[1]{\noindent{\large #1}\vspace{3mm}

}
\newcommand{\address}[1]{\noindent #1\vspace{5mm}

}
\renewcommand{\abstract}[1]{\vspace{19mm}

\noindent{\small{\em Abstract.} #1}\vspace{2mm}

} 
\begin{document}

\vbox{\baselineskip=12pt
\rightline{\small PP97--13}
\rightline{\small gr-qc/9608010}}
\vskip -1 truecm 

\title{Multi-Black-Holes in 3D and 4D anti-de Sitter\\[2mm] 
Spacetimes\footnote{talk given at JR96 (Ascona, Switzerland), to be
published in Helv.~Phys.~Acta}}
\authors{Dieter R. Brill}
\address{Department of Physics, University of Maryland\\
College Park, MD 20742, USA\footnote{e-mail:brill@umdhep.umd.edu}}
\abstract{The (single) black hole solutions of Ba\~nados, Teitelboim and 
Zanelli (BTZ) in 2+1 dimensional anti-de Sitter space are generalized to
an arbitrary number $n$ of such black holes. The resulting multi-black-hole 
(MBH) spacetime is locally isometric to anti-de Sitter space, and globally
it  is obtained from the latter as a quotient space by means of suitable 
identifications. The MBH spacetime has $n$ asymptotically anti-de Sitter  
exterior regions, each of which has the geometry of a single BTZ black hole.
These exterior regions are separated by $n$ horizons from a common interior
region. This interior region can be described as a ``closed" universe 
containing $n$ black holes. Similar configurations in 3+1 dimensions, 
with horizons of toroidal and higher genus topologies, are also presented.}
\section{Introduction}
In 3D Einstein's equations determine the full Riemann tensor. In vacuum
(with only a cosmological constant $\Lambda$) the spacetime has constant
curvature, there are no local gravitational degrees of freedom. If one
thinks of black holes as concentrations of curvature it is surprising
that in 2+1 dimensions there are ``black hole" solutions \cite{BTZ}. 
Their black hole properties derive solely from the global structure of the 
spacetime. 

Because there are no local degrees of freedom it is reasonable that there
should also be multi-black-hole solutions. In this contribution I sketch
the construction of such solutions, and mention the corresponding
spacetimes in 3+1 dimensions.

\section{Initial values}
The black hole construction is possible only if the constant curvature is
{\it negative}, $\Lambda = -1/\ell^2$.
The universal covering space of such spaces is (unwrapped) 3D anti-de
Sitter  (adS) space. Its wrapped version can be embedded in 4D flat
space with signature $--++$, 
$$ds^2 = -dU^2 - dT^2 + dx^2 + dy^2 \eqno(1)$$
as a hyperbolic surface,
$$-U^2-T^2+X^2+Y^2 = -\ell^2. \eqno(2)$$
When the induced metric on this surface is expressed in terms of the 
(non-spinning) BTZ \cite{BTZ} coordinates $r,\,\phi,\,t$,
$$\begin{array}{c}X\\T\end{array}=\sqrt{-\ell^2+{r^2\over M}}\begin{array}
{c}\cosh\\ \sinh\end{array}\left({\sqrt{M}\over\ell}t\right), \qquad
\begin{array}{c}U\\Y\end{array}={r\over\sqrt{M}}\begin{array}
{c}\cosh\\ \sinh\end{array}\left(\sqrt{M}\phi\right) \eqno(3)$$
it takes the form
$$ds^2 = (-M+{r^2\over\ell^2})dt^2 + {dr^2\over-M+{r^2\over\ell^2}}+ 
r^2d\phi^2, \eqno(4)$$
which is similar to that of the Schwarzschild black hole metric. It also
has many similar global properties (as expressed, for example, in a Penrose 
diagram).

The surface $t=$ const is totally geodesic, and therefore its intrinsic
2D geometry is also of constant, negative curvature. The universal covering
space of such constant curvature spaces is 2D hyperbolic space, {\bf
H}$^2$,  which can be represented as the Poincar\'e disk (or as the
Poincar\'e upper half space.) The disk can be obtained by {\em
stereographic projection}
\cite{BS} of the $T=0$ subspace of the surface of Eq (2) on the plane 
$U=-\ell$ from a projection center at $X=Y=0$, $U=\ell$.

The lines $\phi=$ const are geodesics, and are therefore
represented in the Poincar\'e model as circular arcs perpendicular to the
boundary curve (circumference in the case of the disk model). We cut the
$t=0$ initial state of the BTZ black hole along $\phi=0$ and $\phi=\pi$,
obtaining two congruent strips in the Poincar\'e model, one of which is
shown in Figure 1a. To reassemble the BTZ initial state we take two copies 
of this figure and join them together along the boundaries $\phi=0$ and 
$\phi=\pi$. We call this procedure of joining two copies of a region along 
geodesic boundaries ``doubling" of the figure. (That the two copies fit 
together smoothly follows from the identity of the intrinsic geometry of
the  boundaries, and vanishing of their extrinsic curvature, since they are 
geodesics.) The region between any pair of non-intersecting 
geodesics on the Poincar\'e disk, when doubled, yields a BTZ black hole of 
finite mass parameter $M$. If the geodesics meet on the ``limit circle" 
(at infinity), $M$ vanishes; otherwise any such region can be brought in
the symmetrical position of Fig.~1a by means of an isometry of the
Poincar\'e disk. In that position it is obvious that the horizon occurs at
the minimal geodesic connecting the two boundary geodesics, and $M$ is
determined by the length of that geodesic compared to $\ell$.
(Alternatively, $M$ measures the angle that ``infinity" subtends at the
horizon.)

By the same method one can obtain multi-black-hole (MBH) initial states,
with several asymptotically adS regions \cite{B}. Figure 1b shows a strip
which, when doubled, gives an initial state with three asymptotic regions,
which we call a 3-black-hole state. (Figure 1c shows the same geometry
in the upper half plane model.) By identifying two of the horizons, for 
example h$_2$ and h$_3$, we get a ``wormhole-type" geometry with only
one asymptotic region. By identifying horizons by pairs in an originally
4-black-hole geometry we obtain a compact, negative curvature geometry.

\bigskip\bigskip

{\footnotesize
\unitlength 1.50mm
\linethickness{0.4pt}
\begin{picture}(107.00,27.00)(3.5,0)
\bezier{4}(5.00,15.00)(5.11,17.11)(5.67,18.67)
\bezier{4}(5.67,18.67)(6.56,20.78)(7.89,21.89)
\bezier{4}(7.89,21.89)(9.33,23.56)(11.33,24.22)
\bezier{4}(11.33,24.22)(12.89,24.89)(15.00,25.00)
\bezier{4}(15.00,25.00)(17.11,24.89)(18.78,24.33)
\bezier{4}(18.78,24.33)(20.56,23.44)(22.00,22.00)
\bezier{4}(22.00,22.00)(23.78,20.44)(24.22,18.78)
\bezier{4}(24.22,18.78)(25.00,17.11)(25.00,15.00)
\bezier{4}(5.00,15.00)(5.11,12.89)(5.67,11.33)
\bezier{4}(5.67,11.33)(6.56,9.22)(7.89,8.11)
\bezier{4}(7.89,8.11)(9.33,6.44)(11.33,5.78)
\bezier{4}(11.33,5.78)(12.89,5.11)(15.00,5.00)
\bezier{4}(15.00,5.00)(17.11,5.11)(18.78,5.67)
\bezier{4}(18.78,5.67)(20.56,6.56)(22.00,8.00)
\bezier{4}(22.00,8.00)(23.78,9.56)(24.22,11.22)
\bezier{4}(24.22,11.22)(25.00,12.89)(25.00,15.00)
\put(24.00,8.50){\makebox(0,0)[lb]{$\phi=0$}}
\put(26.00,15.00){\makebox(0,0)[lc]{$\phi=\pi/2$}}
\put(24.00,21.20){\makebox(0,0)[lc]{$\phi=\pi$}}
\put(15.00,27.00){\makebox(0,0)[cc]{$r=\ell\sqrt{M}$}}
\put(4.00,23.00){\vector(1,-4){1.25}}
\put(4.00,24.00){\makebox(0,0)[cc]{$r=\infty$}}
\bezier{4}(45.00,15.00)(45.11,17.11)(45.67,18.67)
\bezier{4}(45.67,18.67)(46.56,20.78)(47.89,21.89)
\bezier{4}(47.89,21.89)(49.33,23.56)(51.33,24.22)
\bezier{4}(51.33,24.22)(52.89,24.89)(55.00,25.00)
\bezier{4}(55.00,25.00)(57.11,24.89)(58.78,24.33)
\bezier{4}(58.78,24.33)(60.56,23.44)(62.00,22.00)
\bezier{4}(62.00,22.00)(63.78,20.44)(64.22,18.78)
\bezier{4}(64.22,18.78)(65.00,17.11)(65.00,15.00)
\bezier{4}(45.00,15.00)(45.11,12.89)(45.67,11.33)
\bezier{4}(45.67,11.33)(46.56,9.22)(47.89,8.11)
\bezier{4}(47.89,8.11)(49.33,6.44)(51.33,5.78)
\bezier{4}(51.33,5.78)(52.89,5.11)(55.00,5.00)
\bezier{4}(55.00,5.00)(57.11,5.11)(58.78,5.67)
\bezier{4}(58.78,5.67)(60.56,6.56)(62.00,8.00)
\bezier{4}(62.00,8.00)(63.78,9.56)(64.22,11.22)
\bezier{4}(64.22,11.22)(65.00,12.89)(65.00,15.00)
\thicklines
\put(5.00,15.00){\line(1,0){20.00}}
\bezier{36}(63.58,20.33)(59.58,18.17)(55.00,18.25)
\bezier{20}(64.33,18.42)(62.00,17.05)(62.00,15.00)
\bezier{36}(46.42,20.33)(50.42,18.17)(55.00,18.25)
\bezier{36}(63.58,9.67)(59.58,11.83)(55.00,11.75)
\bezier{20}(64.33,11.58)(62.00,12.95)(62.00,15.00)
\bezier{36}(46.42,9.67)(50.42,11.83)(55.00,11.75)
\thinlines
\put(55.00,18.17){\line(0,-1){6.33}}
\bezier{20}(60.17,18.92)(60.92,17.17)(62.33,16.42)
\bezier{20}(60.17,11.08)(60.92,12.83)(62.33,13.58)
\thicklines
\bezier{36}(23.58,20.33)(19.58,18.17)(15.00,18.25)
\bezier{36}(6.42,20.33)(10.42,18.17)(15.00,18.25)
\bezier{36}(23.58,9.67)(19.58,11.83)(15.00,11.75)
\bezier{36}(6.42,9.67)(10.42,11.83)(15.00,11.75)
\put(15.00,18.17){\line(0,-1){6.33}}
\bezier{32}(10.50,18.67)(9.50,15.00)(10.50,11.33)
\bezier{32}(19.50,18.67)(20.50,15.00)(19.50,11.33)
\thinlines
\put(15.00,25.58){\vector(0,-1){6.83}}
\put(15.00,2.00){\makebox(0,0)[cc]{(a)}}
\put(55.00,2.00){\makebox(0,0)[cc]{(b)}}
\put(95.00,2.00){\makebox(0,0)[cc]{(c)}}
\put(55.20,15.00){\makebox(0,0)[rc]{h$_1$}}
\put(61.50,16.50){\makebox(0,0)[rb]{h$_2$}}
\put(61.00,13.00){\makebox(0,0)[rc]{h$_3$}}
\bezier{20}(83.00,15)(95,15)(107,15)
\thicklines
\bezier{8}(91.90,15.11)(91.93,15.76)(92.11,16.25)
\bezier{8}(92.11,16.25)(92.38,16.90)(92.80,17.25)
\bezier{8}(92.80,17.25)(93.24,17.76)(93.86,17.97)
\bezier{8}(93.86,17.97)(94.35,18.18)(95.00,18.21)
\bezier{8}(95.00,18.21)(95.65,18.18)(96.17,18.00)
\bezier{8}(96.17,18.00)(96.72,17.73)(97.17,17.28)
\bezier{8}(97.17,17.28)(97.72,16.80)(97.86,16.28)
\bezier{8}(97.86,16.28)(98.10,15.76)(98.10,15.11)
\bezier{16}(85.00,15.11)(85.07,17.04)(85.67,18.78)
\bezier{16}(85.67,18.78)(86.53,20.78)(88.00,22.24)
\bezier{16}(88.00,22.24)(89.40,23.71)(91.33,24.38)
\bezier{16}(91.33,24.38)(93.00,25.11)(95.00,25.11)
\bezier{16}(105.00,15.11)(104.93,17.04)(104.33,18.78)
\bezier{16}(104.33,18.78)(103.47,20.78)(102.00,22.24)
\bezier{16}(102.00,22.24)(100.60,23.71)(98.67,24.38)
\bezier{16}(98.67,24.38)(97.00,25.11)(95.00,25.11)
\put(101.5,15){\oval(4.75,4.75)[t]}
\thinlines
\put(95,25){\line(0,-1){7}}
\put(94.9,22.00){\makebox(0,0)[rc]{h$_1$}}
\end{picture}}

\vskip-5pt
{\small \noindent
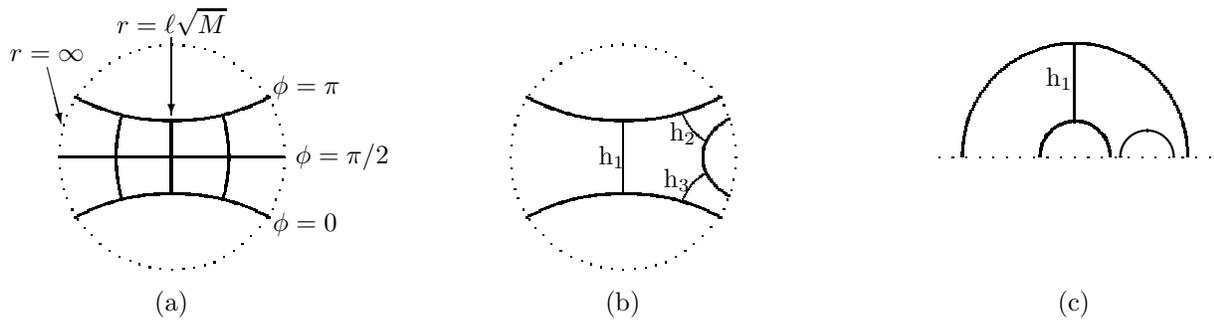
Fig.~1. Construction of MBH initial values, shown in the Poincar\'e
model of {\bf H}$^2$. In this model geodesics are arcs of circles
perpendicular to the boundary (dotted), which represents infinity.\\
(a) In the disk model the ``single" BTZ black hole's initial state is the 
region in which some BTZ coordinate lines are shown (thicker lines). 
This region is to be doubled, as described in the text. The horizon is the 
vertical geodesic segment at $r=\ell\sqrt{M}$.\\ 
(b) Initial state of a ``3-black-hole" configuration, with three 
asymptotically adS regions reaching out to infinity. Each of these exterior
regions is separated from the common interior region by a horizon h
(thinner arcs), the shortest geodesic segment between a pair of
non-intersecting  geodesic boundaries (thicker arcs). The length
$\pi\ell\sqrt{M_i}$ of the  horizon h$_i$ measures the the mass $M_i$
associated with the $i^{\rm th}$  exterior. (In this figure the lengths of
the three horizons, as determined by  the Poincar\'e metric, and hence the
three masses, are actually equal).\\ (c) The same geometry as in (b) in
the upper half-plane model. Only one  of the horizons is shown.}
\section{Time Development}
The left half of Fig.~1b is identical to that of Fig.~1a, hence the 
time development in its domain of dependence will also be the same. But
that  domain is the region outside the spacetime horizon corresponding to
$h_1$; thus to an observer in an exterior region outside of one of the 
horizons h$_i$ the future spacetime geometry is indistinguishable from that 
of a single BTZ black hole. Because the past of such observers includes
more than the exterior's past domain of dependence, they can be aware of
the difference from a single BTZ black hole, for example via light from the
``white hole" singularity.

The region between the horizons of a MBH initial state has finite area;
except for the openings at the horizons it is a closed universe model.
Because it is locally homogeneous, its time development in time-orthonormal
coordinates will  amount simply to an overall scale change, described for
example by the Raychaudhuri equation \cite{BS}. The scale factor decreases
to zero in the  finite proper time $\pi\ell/2$, at which time a
non-Hausdorff singularity,  similar to that of the BTZ black hole
\cite{NH}, is reached.
\section{Analogous Configurations in 3+1 Dimensions}
The BTZ idea can be generalized to 4D spacetimes in various ways \cite{Be}. 
The following metrics, found in collaboration with Dr.~J.~Louko and
obtained by analytic continuation and re-scaling of the Schwarzschild-de
Sitter metric 
\cite{C}, satisfy the vacuum Einstein equations with cosmological constant 
$\Lambda$:
$$ds^2 = -Fdt^2 + {dr^2\over F} + 
r^2\left(\begin{array}{c}d\theta^2 +\cosh^2\sqrt{M}\theta\, d\phi^2\\ 
d\theta^2 +\theta^2 d\phi^2\end{array}\right)
\quad {\rm with} \, F = \begin{array}{c}-M-{2m\over r}-\Lambda r^2 \quad
(M\neq 0)\\
-{2m\over r}-\Lambda r^2\qquad (M=0).\end{array} $$
When $\Lambda =-1/\ell^2$ is negative this has the appropriate signature
at large $r$ and becomes the adS metric there, and when $m=0$ and
$\theta=0$ it becomes the BTZ metric (4). However, to complete the analogy,
the surfaces  of constant $r$ and $t$ should be compactified. In the case
$M=0$ these 2D  surfaces are flat, so a torus compactification is possible.
If
$M\neq 0$ these surfaces have constant negative curvature, and can also be
compactified,  for example as described at the end of Section 2. When the
Schwarzschild-type  mass parameter $m$ vanishes, joining two or more of
these in their anti-de  Sitter regions yields MBH geometries. Simple
joining is not possible when  $m \neq 0$ because the curvature is then no
longer constant. Instead,  the single black hole with $m \neq 0$ has a
curvature singularity at $r=0$,  like the Schwarzschild geometry.

\end{document}